\documentclass[12pt]{article}

\usepackage{amsfonts,amssymb,amsmath}
\usepackage[dvips]{epsfig}
\begin{document}
\title{\bf Requirement of system-reservoir bound states for entanglement protection}
\author{ N. Behzadi $^{a}$
\thanks{E-mail:n.behzadi@tabrizu.ac.ir}  ,
B. Ahansaz $^{b}$,
E. Faizi $^{b}$ and
H. Kasani $^{c}$
\\ $^a${\small Research Institute for Fundamental Sciences, University of Tabriz, Tabriz, Iran,}
\\ $^b${\small Physics Department, Azarbaijan Shahid Madani University, Tabriz, Iran,}
\\ $^c${\small Physics Department, University of Mohaghegh Ardabili, Ardabil, Iran.}}\maketitle

\begin{abstract}
\noindent
In this work, a genuine mechanism of entanglement protection of a two-qubit system interacting with a dissipative common reservoir is investigated. Based on the generating of bound state for the system-reservoir, we show that stronger bound state in the energy spectrum can be created by adding other non-interacting qubits into the reservoir. In the next step, it is found that obtaining higher degrees of boundedness in the energy spectrum leads to better protection of two-qubit entanglement against the dissipative noises. Also, it is figured out that the formation of bound state not only exclusively determines the long time entanglement protection, irrespective to the Markovian and non-Markovian dynamics, but also performs the task for reservoirs with different spectral densities.
\\
\\
{\bf PACS Nos:}
\end{abstract}

\section{Introduction}
Quantum entanglement, one of the wonderful aspects of quantum mechanics which has no classical counterpart,
has been considered as a main resource in understanding and development of quantum information processing protocols ~\cite{Nielsen}.
However, since quantum entanglement is so fragile due to decoherence effect, it inevitably undergoes either asymptotic decay or sudden death processes ~\cite{Hornberger, Merali}.
So in this regard, unavoidable interaction between any real quantum system and its surrounding environment alters the quantum system and consequently disentanglement occurs. This is apparently a disadvantageous procedure relative to the applications of quantum information processing tasks.
Therefore, it is very important to find out a mechanism whose advantage leads to an effective long-time entanglement preservation. So far, a lot of researches have been devoted to entanglement manipulation and protection, such as quantum Zeno effect (QZE) ~\cite{Maniscalco, Mundarain, Rossi, Hou} and
detuning modulation ~\cite{Ba, Xiao}.

It was found that the coupling between an excited atom and electromagnetic vacuum reservoir leads atom-photon bound state ~\cite{John2} which in turns exhibits a fractional steady-state atomic population on the excited state known as populating trapping ~\cite{Lambropoulos}. On the other hand, entanglement trapping due to structured environment ~\cite{Bellomo} is a direct consequence of populating trapping. Based on these finding, it was proposed a method for entanglement protection of two qubits in two uncorrelated reservoirs using the non-Markovian effect along with creating bound state for the system-environment through the manipulating of the spectral densities ~\cite{Tong}.

In this paper, we give an actual mechanism for protection of two-qubit entanglement exclusively on the basis of formation of bound state for the two-qubit system and its common reservoir using additional qubits. In this approach without manipulating the spectral density of the environment, system-reservoir bound state can be generated by adding other non-interacting qubits into the reservoir. It is interesting to note that creating bound state in the energy spectrum of the total system with higher degree of boundedness is provided by inserting more additional qubits. As will be figured out, higher degree of boundedness gives the better protection of entanglement in long time limit irrespective to Markovian or non-Markovian dynamics. We also examine the performance of the approach for several different spectral densities and observe that the protection process only depends on the existence of bound state and completely independent from structure of environment.

This paper is organized as follows. In Sec. II, the model of $N$ non-interacting qubits in a common zero-temperature reservoir is introduced. We derive the condition for the formation of bound state in the spectrum of total system. In Sec. III, the entanglement dynamics of two-qubit system is studied by considering different models for the reservoir structure. Finally, a brief conclusions are given at the end of the paper.

\section{The model}
In this section, we consider a system of $N$ non-interacting qubits (two-level atoms) immersed in a common zero-temperature thermal reservoir.
The Hamiltonian $\hat{H}$ of the system can be written as ($\hbar=1$)
\begin{eqnarray}
\hat{H}=\hat{H}_{0}+\hat{H}_{I},
\end{eqnarray}
where $\hat{H}_{0}$ is the free Hamiltonian and $\hat{H}_{I}$ describes the interaction terms,
\begin{eqnarray}
\hat{H}_{0}=\omega_{0} \sum_{l=1}^{N} \hat{\sigma}^{+}_{l} \hat{\sigma}^{-}_{l}+\sum_{k} \omega_{k} \hat{b_{k}}^{\dagger} \hat{b_{k}},
\end{eqnarray}

\begin{eqnarray}
\hat{H}_{I}=\sum_{l=1}^{N} \sum_{k} g_{k} \hat{b_{k}} \hat{\sigma}_{l}^{+}+g_{k}^{*} \hat{b_{k}^{\dagger}} \hat{\sigma}_{l}^{-}.
\end{eqnarray}
In Eqs. (2) and (3), $\hat{\sigma}_{l}^{+}$ $(\hat{\sigma}_{l}^{-})$ is the raising (lowering) operator of the $l^{th}$ qubit with transition frequency $\omega_{0}$
and $\hat{b_{k}}$ ($\hat{b_{k}}^{\dagger}$) is the annihilation (creation) operator of the $k^{th}$ field mode with frequency $\omega_{k}$.
Also, the strength of coupling between the $l^{th}$ qubit and the $k^{th}$ field mode is represented by $g_{k}$.
The spectrum of the Hamiltonian can be obtained by solving the following eigenvalue equation
\begin{eqnarray}
 \hat{H}(t) |\psi(t)\rangle=E |\psi(t)\rangle.
\end{eqnarray}
Since the total Hamiltonian commutes with the number of excitations (i. e. $[(\sum_{l=1}^{N} {\hat{\sigma}_{l}^{+} \hat{\sigma}_{l}^{-}}+\sum_{k} \hat{b_{k}}^{\dagger} \hat{b_{k}}),H]=0$), therefore by considering the single excitation subspace, we have
\begin{eqnarray}
  |\psi(t)\rangle=\sum_{l=1}^{N}C_{l}(t)|l\rangle_{s} |0\rangle_{e} +\sum_{k}C_{k}(t)|0\rangle_{s} |1_{k}\rangle_{e}.
\end{eqnarray}
where $|l\rangle_{s}=|g\rangle^{\bigotimes N}_{l^{th}\equiv e}$, which means that all of the qubits are in their respective ground states $|g\rangle$ except the $l^{th}$ qubit which is in the excited state $|e\rangle$, and  $|0\rangle_{s}=|g\rangle^{\bigotimes N}=|g,g,...,g\rangle$.
Also, we denote $|0\rangle_{e}$ being the vacuum state of the reservoir and $|1_{k}\rangle_{e}$ is the state for which there is only one excitation in the $k$th field mode.
Substituting Eqs. (1) and (5) into Eq. (4),  yields the following set of $N+1$ equations
\begin{eqnarray}
\begin{array}{c}
  \omega_{k} C_{k}(t)+\sum_{l=1}^{N} g_{k}^{*} C_{l}(t)=E C_{k}(t),\\\\
  \omega_{0} C_{1}(t)+\sum_{k} g_{k} C_{k}(t)=E C_{1}(t),\\\\
  \omega_{0} C_{2}(t)+\sum_{k} g_{k} C_{k}(t)=E C_{2}(t),\\
  .\\
  .\\
  .\\
  \omega_{0} C_{N}(t)+\sum_{k} g_{k} C_{k}(t)=E C_{N}(t).
\end{array}
\end{eqnarray}
Obtaining $C_{k}(t)$ from the first equation and substituting it in the rest ones gives the following $N$ integro-differential equations
\begin{eqnarray}
\begin{array}{c}
  (E-\omega_{0}) C_{1}(t)=-\int_{0}^{\infty} \frac{J(\omega) d\omega}{\omega-E} \sum_{l=1}^{N} C_{l}(t),\\\\
  (E-\omega_{0}) C_{2}(t)=-\int_{0}^{\infty} \frac{J(\omega) d\omega}{\omega-E} \sum_{l=1}^{N} C_{l}(t),\\
  .\\
  .\\
  .\\
  (E-\omega_{0}) C_{N}(t)=-\int_{0}^{\infty} \frac{J(\omega) d\omega}{\omega-E} \sum_{l=1}^{N} C_{l}(t).
\end{array}
\end{eqnarray}
Eliminating the coefficients $C_{l}(t)$s reads
\begin{eqnarray}
E=\omega_{0}-N \int_{0}^{\infty} \frac{J(\omega) d\omega}{\omega-E} \equiv y(E).
\end{eqnarray}
Solving Eq. (8) gives the energy spectrum of the total system, i.e.  $N$-qubit system and reservoir, which depends effectively on the spectral density of the reservoir.
Note that there is a solution for Eq. (8) when the functions $y(E)$ and $y(E)=E$ are crossed with each other.
Since $y(E)$ decreases monotonically with the increase of $E$ in the region $E < 0$ and $\mathrm{lim}_{E \rightarrow -\infty} y(E)=\omega_{0}$, so Eq. (8) has an
isolated root in this region. The eigenstate corresponding to this isolated eigenvalue is named the bound state which exists between the $N$-qubit system and its common reservoir.
It can be easily checked out that there is bound state solution for Eq. (8) when it satisfies the condition $y(0) < 0$, otherwise, there is no bound state.
On the other hand, it has only complex solutions when a bound state is not formed.
This means that the corresponding eigenstate experiences decay from the imaginary part of the eigenvalue during the
time evolution and therefore the excited-state population approaches to zero asymptotically.
However, for a bound state the population of the atomic excited state
is constant in time because it has a vanishing decay rate during the time evolution.
In the following, we consider some typical spectral functions such as Lorentzian, sub-Ohmic, Ohmic and super-Ohmic for the structure of the reservoir, and show how the addition of non-interacting qubits generates bound state and affects its quality.

\section{Entanglement dynamics}
To quantify the degree of entanglement between of a two-qubit system, we use concurrence as a measure of entanglement in this way~\cite{Wootterrs}.
The concurrence is defined as
\begin{eqnarray}
  C(\rho)=\mathrm{max} \{0,\sqrt{\lambda_{1}}-\sqrt{\lambda_{2}}-\sqrt{\lambda_{3}}-\sqrt{\lambda_{4}}\},
\end{eqnarray}
in which $\lambda_{i}$'s are the eigenvalues, in decreasing order, of the Hermitian matrix $R=\sqrt{\sqrt{\rho}\tilde{\rho}\sqrt{\rho}}$ with $\tilde{\rho}=(\sigma_{y}^{A}\otimes \sigma_{y}^{B}) \rho^{*}(\sigma_{y}^{A}\otimes \sigma_{y}^{B})$. Here $\rho^{*}$ means the complex conjugation of $\rho$ and $\sigma_{y}$ is the Pauli matrix.
Let us consider $m^{th}$ and $n^{th}$ qubits of the system with $m\neq n$, prepared initially in an EPR-type entangled state as follows
\begin{eqnarray}
  |\phi(0)\rangle_{m,n}=C_{m}(0)|e,g\rangle+C_{n}(0)|g,e\rangle.
\end{eqnarray}
Therefore, at time $t>0$, the concurrence of the two-qubit system prepared initially in (10) is as
\begin{eqnarray}
  C(\rho_{m,n}(t))=2|C_{m}(t) C_{n}(t)|.
\end{eqnarray}
To finding the probability amplitudes $C_{m}(t)$ and $C_{n}(t)$, we use the Schrodinger equation in the interaction picture
\begin{eqnarray}
  i \frac{d}{dt}|\psi(t)\rangle=\hat{H}_{I}(t) |\psi(t)\rangle,
\end{eqnarray}
where the Hamiltonian in this picture is given by $\hat{H}_{I}(t)=e^{i \hat{H}_{0}t} \hat{H}_{I} e^{-i \hat{H}_{0}t}$ and substituting Eq. (5) and $\hat{H}_{I}(t)$ into Eq. (12) gives a system of $N+1$ differential equations as
\begin{eqnarray}
  \dot{C}_{l}(t)=-i \sum_{k} g_{k} C_{k}(t) e^{i(\omega_{0}-\omega_{k})t},\\
  \dot{C}_{k}(t)=-i \sum_{l=1}^{N} g_{k}^{*} C_{l}(t) e^{-i(\omega_{0}-\omega_{k})t},
\end{eqnarray}
where $l=1,2,...,N$. Integrating Eq. (14) and substituting it into the Eq. (13) gives
the following set of $N$ closed integro-differential equation
\begin{eqnarray}
  \frac{dC_{l}(t)}{dt}=-\int_{0}^{t} f(t-t') \sum_{u=1}^{N}  C_{u}(t) dt',
\end{eqnarray}
where the correlation function $f(t-t')$ is related to the spectral density $J(\omega)$ of the reservoir by
\begin{eqnarray}
f(t-t')=\int d\omega J(\omega) e^{i(\omega_{0}-\omega)(t-t')}.
\end{eqnarray}
The solutions of Eq. (15), i.e. $C_{l}(t)$'s, depend on the particular choice of the spectral density of the reservoir, which will be considered in the following subsections.

\subsection{Lorentzian spectral density}
We first consider the Lorentzian spectral density as
\begin{eqnarray}
  J(\omega)=\frac{1}{2\pi} \frac{\gamma_{0} \lambda^{2}}{(\omega-\omega_{0})^2+\lambda^{2}},
\end{eqnarray}
where $\omega_{0}$ is the central frequency of the reservoir equal to the transition frequency of the qubits. The parameter $\lambda$ defines the spectral width of the coupling and $\gamma_{0}$ is
the coupling strength. By substituting $J(\omega)$ into Eq. (16), the correlation function $f(t-t')$ can be determined analytically. Using Laplace transform, the exact solutions of Eq. (15) can be obtained as
\begin{eqnarray}
\begin{array}{c}
  C_{l}(t)=e^{-\lambda t/2}\Big(\mathrm{cosh}{(\frac{Dt}{2})}+\frac{\lambda}{D} \mathrm{sinh}{(\frac{Dt}{2}})\Big) C_{l}(0)+\\\\
  (\frac{(N-1) C_{l}(0)-\sum_{l'\neq l}^{N} C_{l'}(0)}{N})\times \\\\
  \Big(1-e^{-\lambda t/2}\big(\mathrm{cosh}{(\frac{Dt}{2})}+\frac{\lambda}{D} \mathrm{sinh}{(\frac{Dt}{2})}\big)\Big),
\end{array}
\end{eqnarray}
where $D=\sqrt{\lambda^{2}-2\gamma_{0} \lambda N}$, and by considering the point that the dynamics is Markovian in the weak coupling regime $(\gamma_{0}<\frac{\lambda}{2N})$ and non-Markovian in the strong coupling regime $(\gamma_{0}>\frac{\lambda}{2N})$.

For the Lorentzian spectral density which is physically corresponding to a single mode leaky cavity, solutions of Eq. (8) in the bound state region ($E<0$) are given in Fig. $1a$ and Fig. $1c$, corresponding to the Markovian and non-Markovian dynamics respectively. For $N=2$, corresponding to the absence of additional qubits, there is no bound state solution for Eq. (8). Absence of the bound state leads complete decay of two-qubit entanglement for both Markovian and Non-Markovian dynamics as shown in Fig. $1b$ and Fig. $1d$, respectively. For $N=8$, corresponding to the presence of six additional qubits, we have formation of the bound state for Markovian and non-Markovian regimes. In this situation, the two-qubit entanglement is suppressed from death and there is a non-zero steady value for the concurrence. As the number of additional qubits grows (for example, $N=12$) we have a bound state whose boundedness becomes stronger which consequently, irrespective to Markovian and Non-Markovian dynamics, gives a better protection of entanglement.

It is concluded that the generation of bound state for the system-environment through the addition of non-interacting qubits actually leads to the protection of two-qubit entanglement against the dissipative noises irrespective to the point that whether its dynamics is Markovian or non-Markovian.

\subsection{Sub-Ohmic, Ohmic and super-Ohmic spectral density}
To making an extension for confirming the performance of the proposed mechanism, we consider other spectral densities whose unified form is
\begin{eqnarray}
  J(\omega)=\frac{\gamma}{2\pi} \omega_{c}^{1-s} \omega^s e^{-\frac{\omega}{\omega_{c}}}.
\end{eqnarray}
where the parameters $s=1/2$, $1$, $2$, are corresponding to sub-Ohmic, Ohmic and super-Ohmic spectral densities respectively, each of which in turns corresponds to a different physical context. It is known that the sub-Ohmic case corresponds to the type of noise appearing in solid state devices, the Ohmic case corresponds to the charged interstitials and the super-Ohmic case corresponds to a phonon bath ~\cite{Paavola}.
The cut-off frequency is represented by $\omega_{c}$ and $\gamma$ is a dimensionless coupling constant.
Unfortunately, there is no analytical solution for the closed integro-differential equation, i.e. Eq. (15), and so the only recourse to
obtaining the concurrence evolution is numerical methods. Also according to Eq. (8), we obtain numerically the condition of existence of bound state for the sub-Ohmic, Ohmic and super-Ohmic cases to investigate how the availability of the bound state is improved by addition of other non-interacting qubits into
the reservoir. As for the Lorentzian spectral density, for each of the mentioned situations, formation of bound state has a determinative role in protection of entanglement against the dissipation. Fig. 2$a$, Fig. 2$c$ and Fig. 2$e$ indicate the fruitful effects of additional qubits in creating system-environment bound state for sub-Ohmic, Ohmic and super-Ohmic spectral densities respectively. Also, it is observed that in the absence of additional qubits ($N=2$) there is no bound state in the system environment spectrum corresponding to the sub-Ohmic, Ohmic and super-Ohmic spectral densities. Therefore, under this condition, protection of entanglement is completely failed (see Fig. 2$b$, Fig. 2$d$ and Fig. 2$f$).

On the other hand, in the attendance of additional qubits into the reservoir ($N=8$), formation of bound state relative to the respective spectral densities is taken place which in turns provides the protection of entanglement from sudden death. As observed for the Lorentzian spectral density in the previous subsection, entering more additional qubits into the reservoir creates bound state with higher degree of boundedness which, in this regard, gives better protection of entanglement again.

Consequently, irrespective to the structure of environment, existence of system environment bound state completely determines the protection process of entanglement in long time limit from dissipation.

At the end, to give a reasonable physical explanation for our results in this paper, we remember that the state of whole system in Eq. (5)
can be expanded in terms of eigenstates of the system-reservoir Hamiltonian as
\begin{eqnarray}
  |\psi(t)\rangle=D_{\mathrm{BS}}e^{-iE_{\mathrm{BS}}t}|\varphi_{\mathrm{BS}}\rangle+\sum_{j\in \mathrm{CB}}D_{\mathrm{CB}}^{j} e^{-iE_{\mathrm{CB}}^{j}t}|\varphi_{\mathrm{CB}}^{j}\rangle.
\end{eqnarray}
where $|\varphi_{\mathrm{BS}}\rangle$ is the potentially formed system-reservoir bound state as an isolated eigenstate with eigenenergy $E_{\mathrm{BS}}$,
$|\varphi_{\mathrm{CB}}\rangle$'s are eigenstates corresponding to the (quasi)continuous energy band, $D_{\mathrm{BS}}=\langle \varphi_{\mathrm{BS}}|\psi(0)\rangle$ and $D_{\mathrm{CB}}^{j}=\langle \varphi_{\mathrm{CB}}^{j}|\psi(0)\rangle$. All of the population terms in the summation of Eq. (20) tend to vanish due to the out-of-phase interference contributed by the (quasi)continuous spectrum.
Therefore, only the first term in Eq. (20) survives in the long-time limit and plays an important role in controlling the evolution of the whole system.
In the absence of additional qubits ($N=2$), the bound state $|\varphi_{\mathrm{BS}}\rangle$ is not formed (see Figs. 1(a,c) and 2(a,c,e)),
so according to Eq. (20), no entanglement can be observed between the considered two qubits ($m^{th}$ and $n^{th}$ qubits) in the long-time limit.
But, inserting the additional qubits into the reservoir ($N=8,12$), leads to formation of the respective bound states
which can be entangled with respect to the mentioned two qubits.
As a result, the time evolution of the two-qubit entanglement ultimately reaches to a non-zero steady value.

\section{Conclusions}

We investigated a mechanism for two-qubit entanglement protection from dissipation caused by a common reservoir on the basis of creating bound states for the system and environment using non-interacting additional qubits. As observed for the reservoir with Lorentzian spectral density, irrespective to the dynamics of the two-qubit system from Markovian and non-Markovian point of views, stronger bound state leads to better protection of entanglement. In the next step, we examined the procedure for other reservoirs with different structures such as sub-Ohmic, Ohmic and supe-Ohmic. It was concluded that the formation of bound state has exclusively a determinative role in the long time entanglement protection irrespective to the structure of the reservoir. Finally, inspired by this mechanism, other methods for manipulating of bound states for protection of entanglement could be introduced in future.
For example, entanglement protection of a two-qudit system by using this approach can be regarded as the subject of our future research.
As a special case, for preservation of entanglement between two qutrits, it can be investigated that whether the addition of qubits leads to the desired result or the qutrits.

\newpage
Fig. 1. (a, c) Diagrammatic solutions of Eq. (8), and (b, d) concurrence dynamics of a two-qubit system
with $N = 2$ (solid line), $N = 8$ (dashed line) and $N = 12$ (dotted line) for a Lorentizan spectral density.
The initial state in Eq. (10) is determined by $C_{m}(0)=C_{n}(0)=1/\sqrt{2}$.
We assume that panels (a, b) are plotted in the Markovian regime, $\lambda=15$ (in units of $\omega_{0}$) and $\gamma_{0}=0.2$ (in units of $\omega_{0}$), and panels (c, d)
in the non-Markovian regime, $\lambda=0.5$ (in units of $\omega_{0}$) and $\gamma_{0}=1$ (in units of $\omega_{0}$).

\begin{figure}
        \qquad \qquad\qquad\qquad \qquad a \qquad\qquad \quad\qquad\qquad\qquad\qquad\qquad\qquad b\\{
        \includegraphics[width=3in]{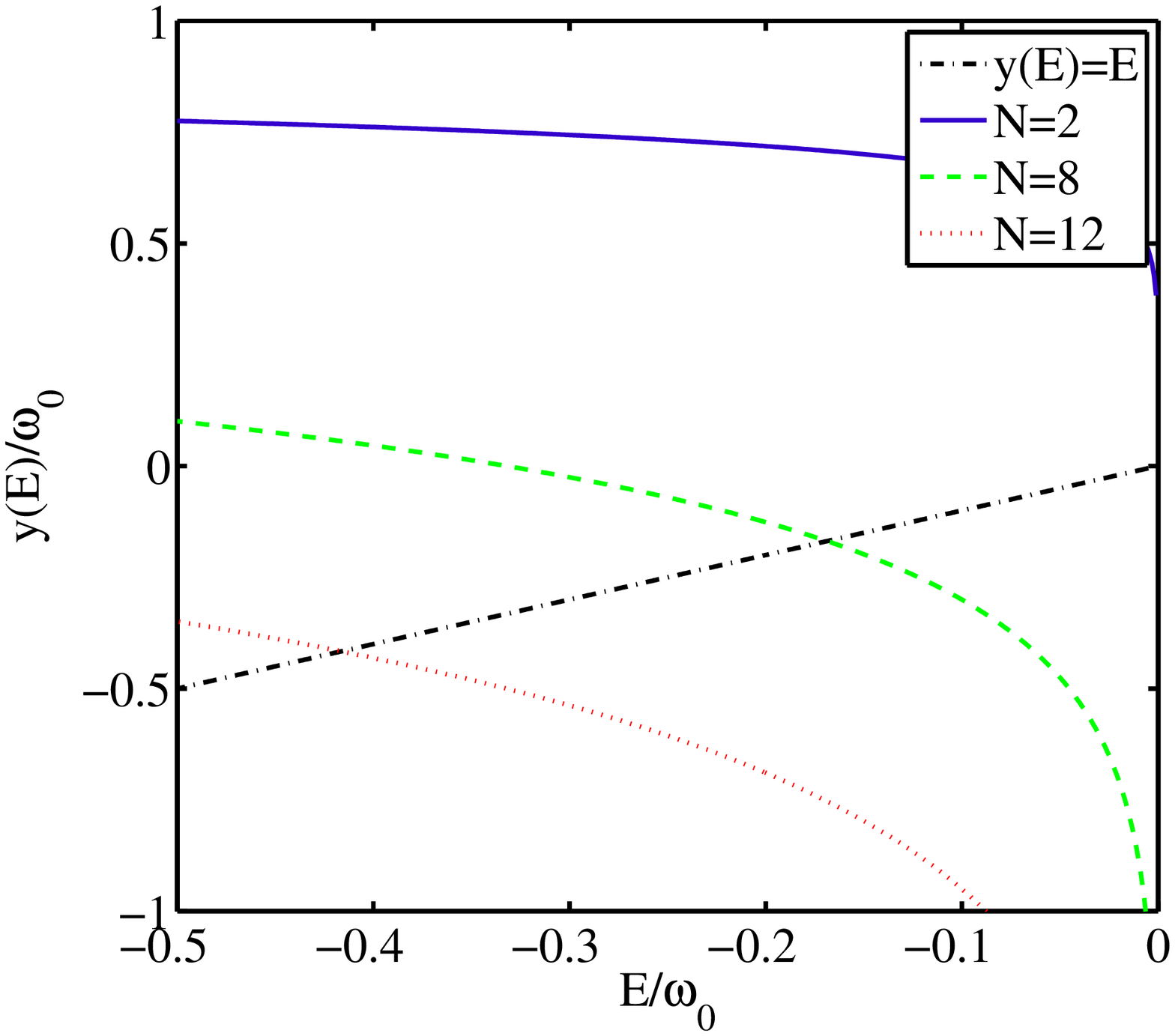}
        \label{fig:first_sub}
    }{
        \includegraphics[width=3in]{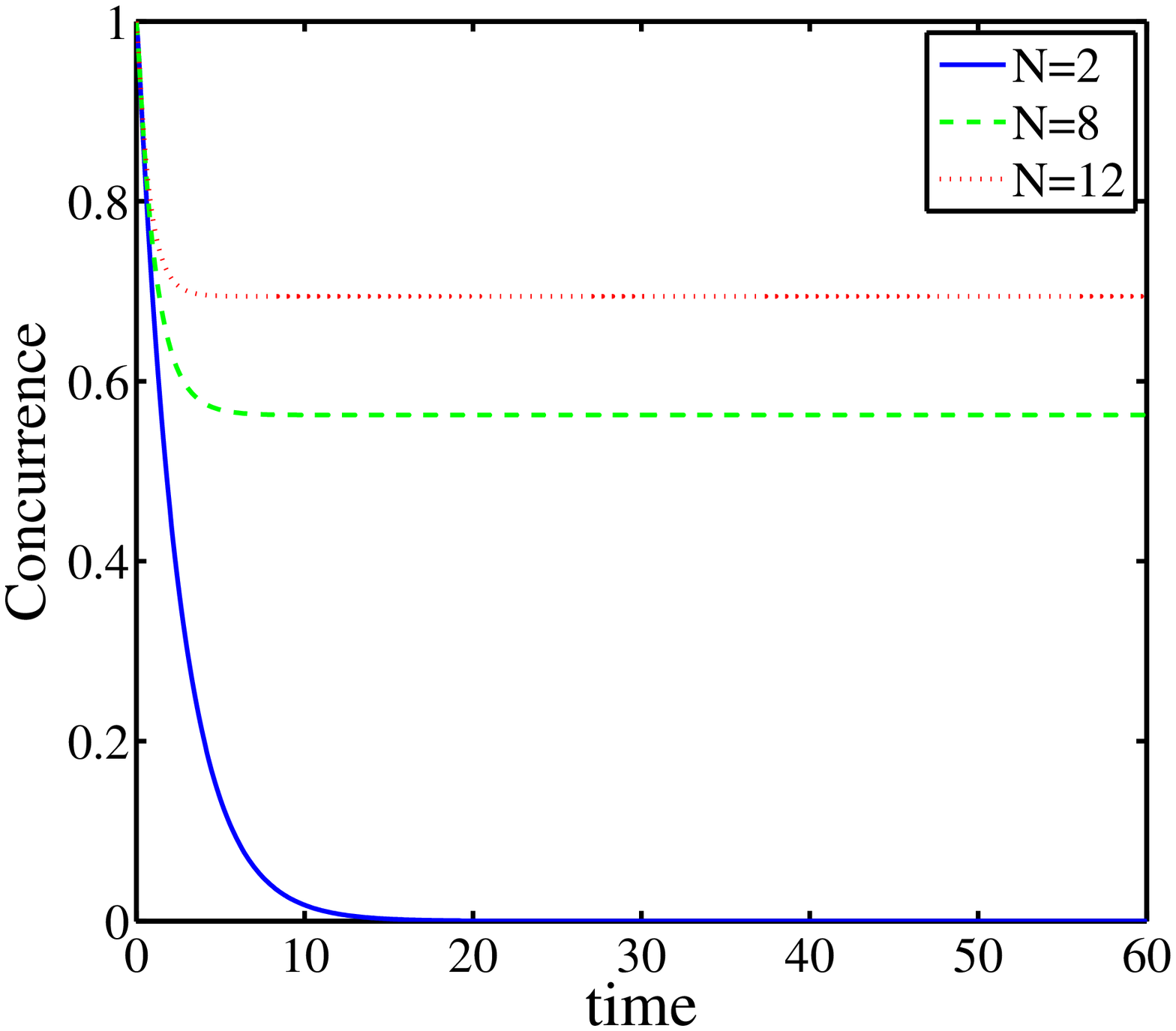}
        \label{fig:second_sub}
    }\\ \par \quad \quad\qquad\qquad\qquad \qquad c \qquad \qquad\qquad\qquad\qquad\quad\qquad\qquad\qquad d\\{
        \includegraphics[width=3in]{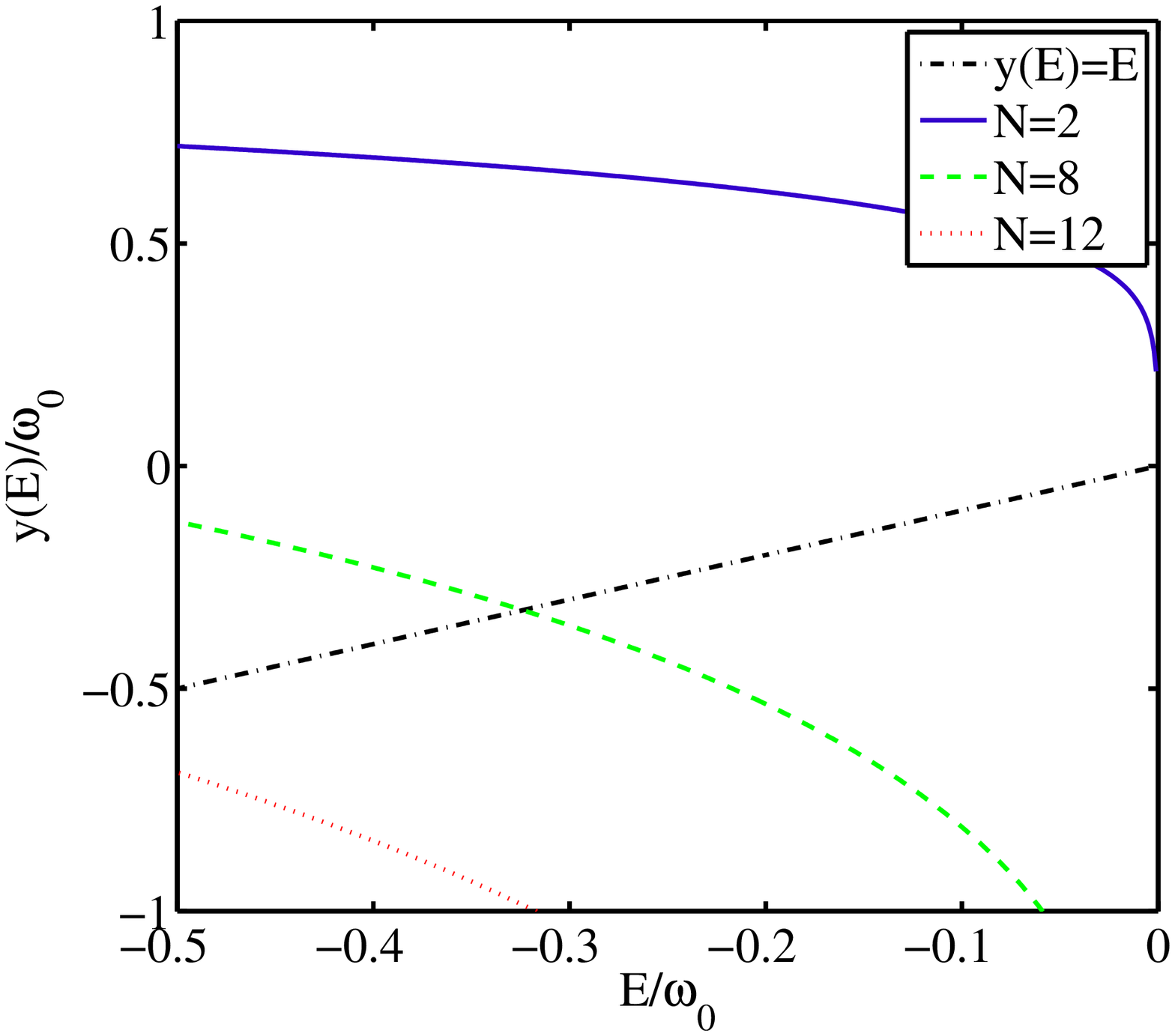}
        \label{fig:first_sub}
    }{
        \includegraphics[width=3in]{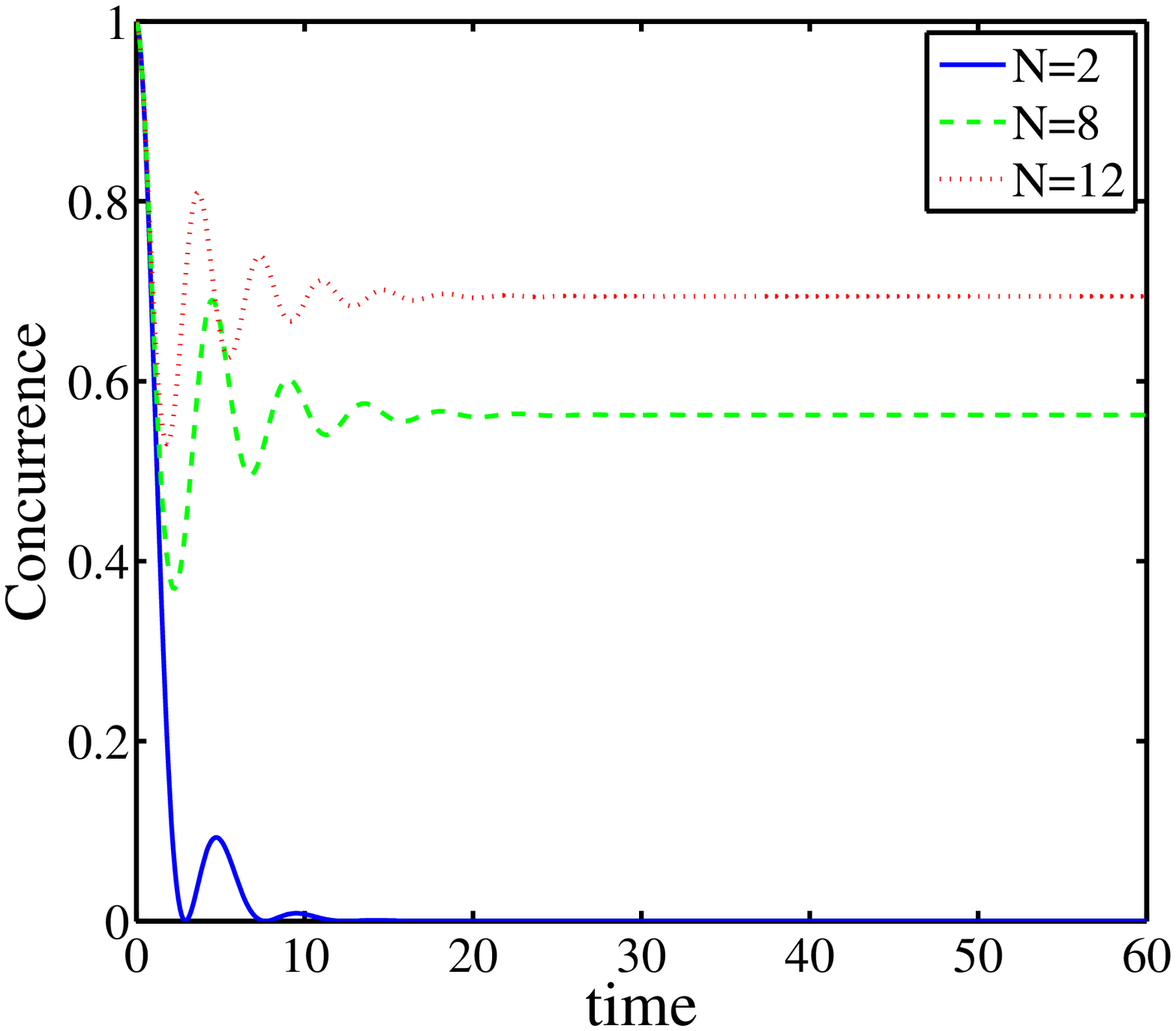}
        \label{fig:second_sub}
    }
    \caption{}
    \end{figure}

\newpage
Fig. 2. (a, c, e) Diagrammatic solutions of Eq. (8), and (b, d, f) concurrence dynamics of a two-qubit system
with $N = 2$ (solid line), $N = 8$ (dashed line) and $N = 12$ (dotted line)
for $(a, b)$ sub-Ohmic $(s = 1/2)$, $(c, d)$ Ohmic $(s = 1)$ and $(e, f)$ super-Ohmic $(s = 2)$ spectral densities with fixed parameters $\omega_{c}=1$ (in units of $\omega_{0}$), $\gamma=1$ (in units of $\omega_{0}$). The initial state in Eq. (10) is determined by $C_{m}(0)=C_{n}(0)=1/\sqrt{2}$.

\begin{figure}
        \qquad \qquad\qquad\qquad \qquad a \qquad\qquad \quad\qquad\qquad\qquad\qquad\qquad\qquad b\\{
        \includegraphics[width=3in]{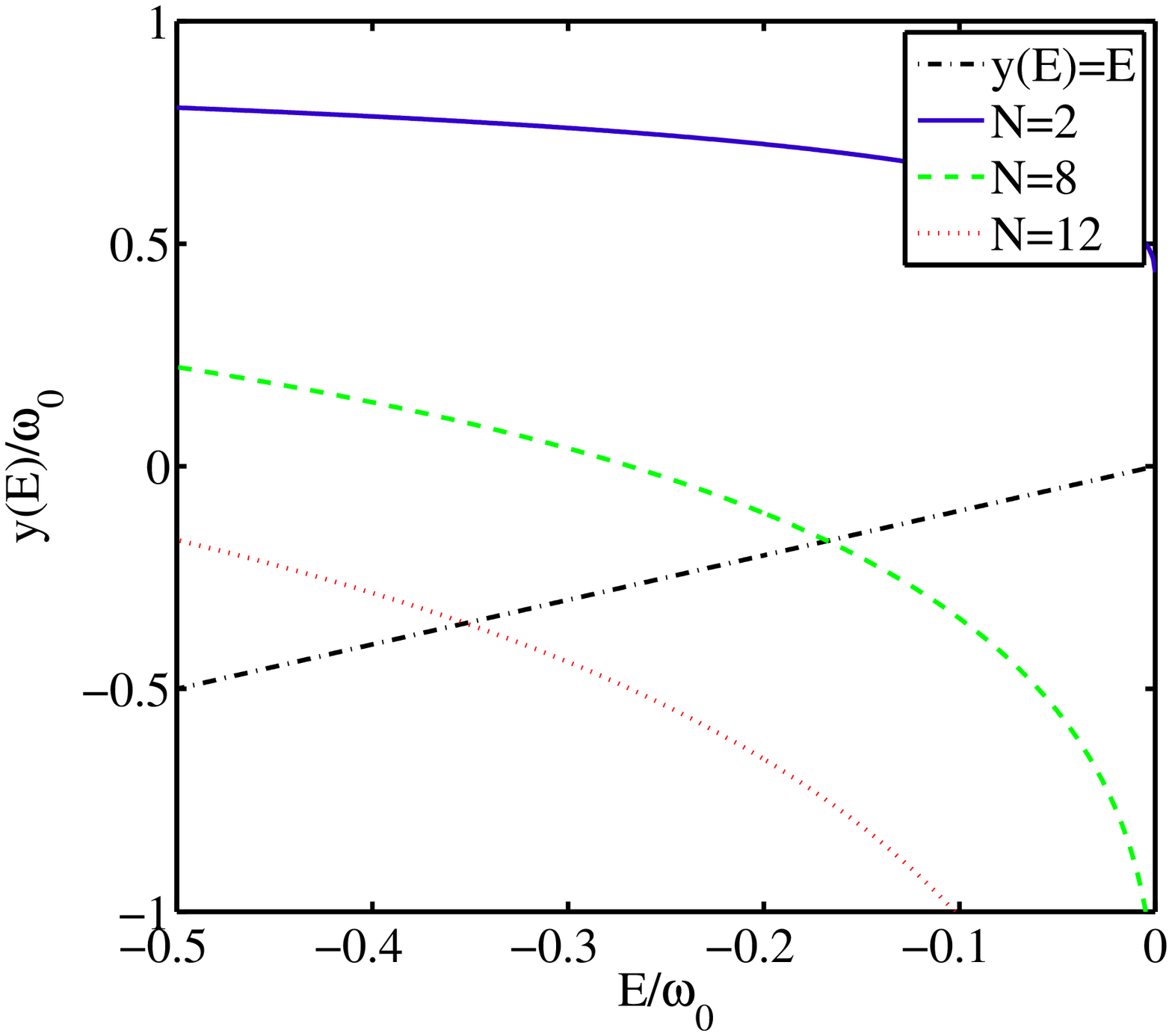}
        \label{fig:first_sub}
    }{
        \includegraphics[width=3in]{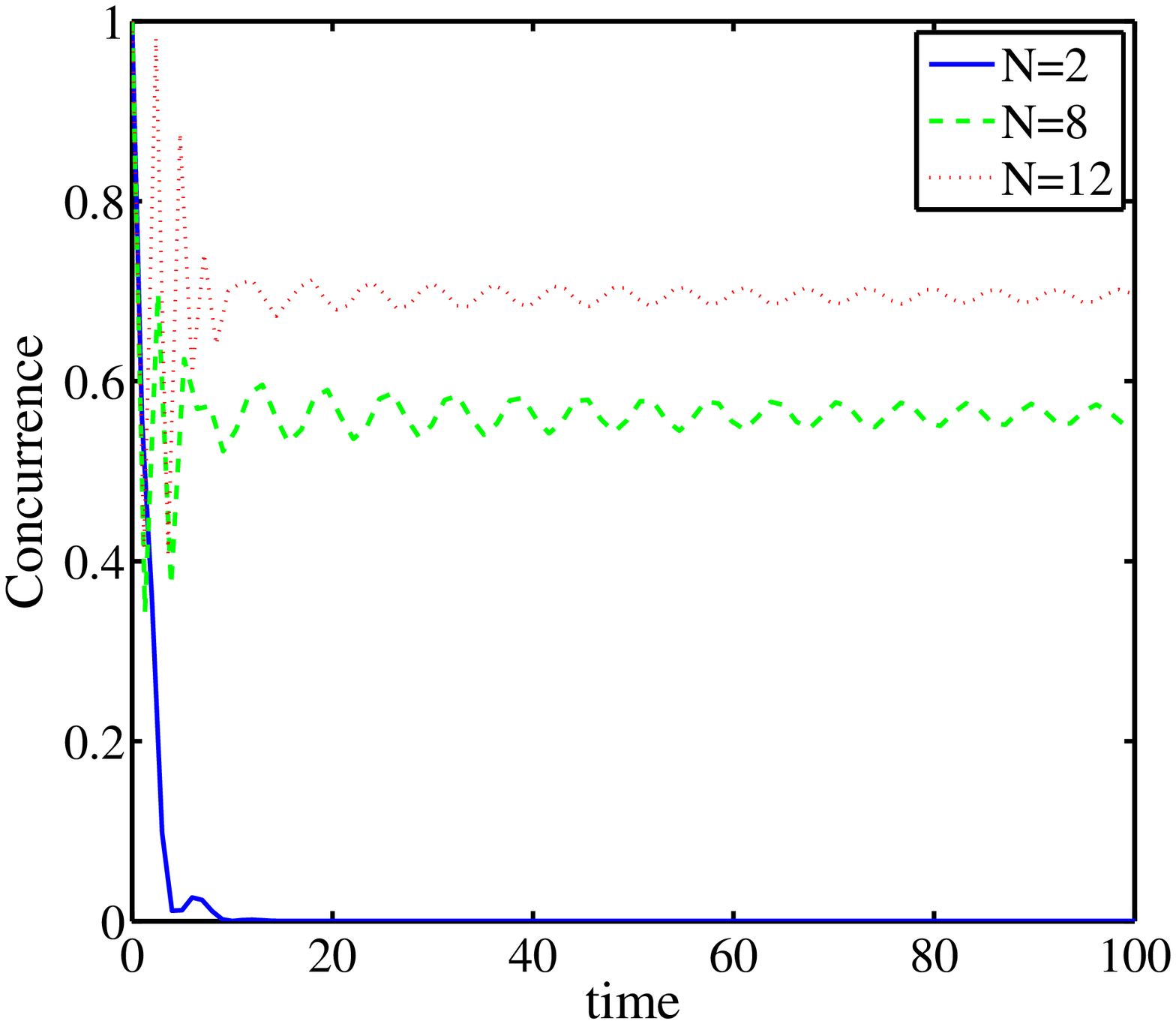}
        \label{fig:second_sub}
    }\\ \par \quad \quad\qquad\qquad\qquad \qquad c \qquad \qquad\qquad\qquad\qquad\quad\qquad\qquad\qquad d\\{
        \includegraphics[width=3in]{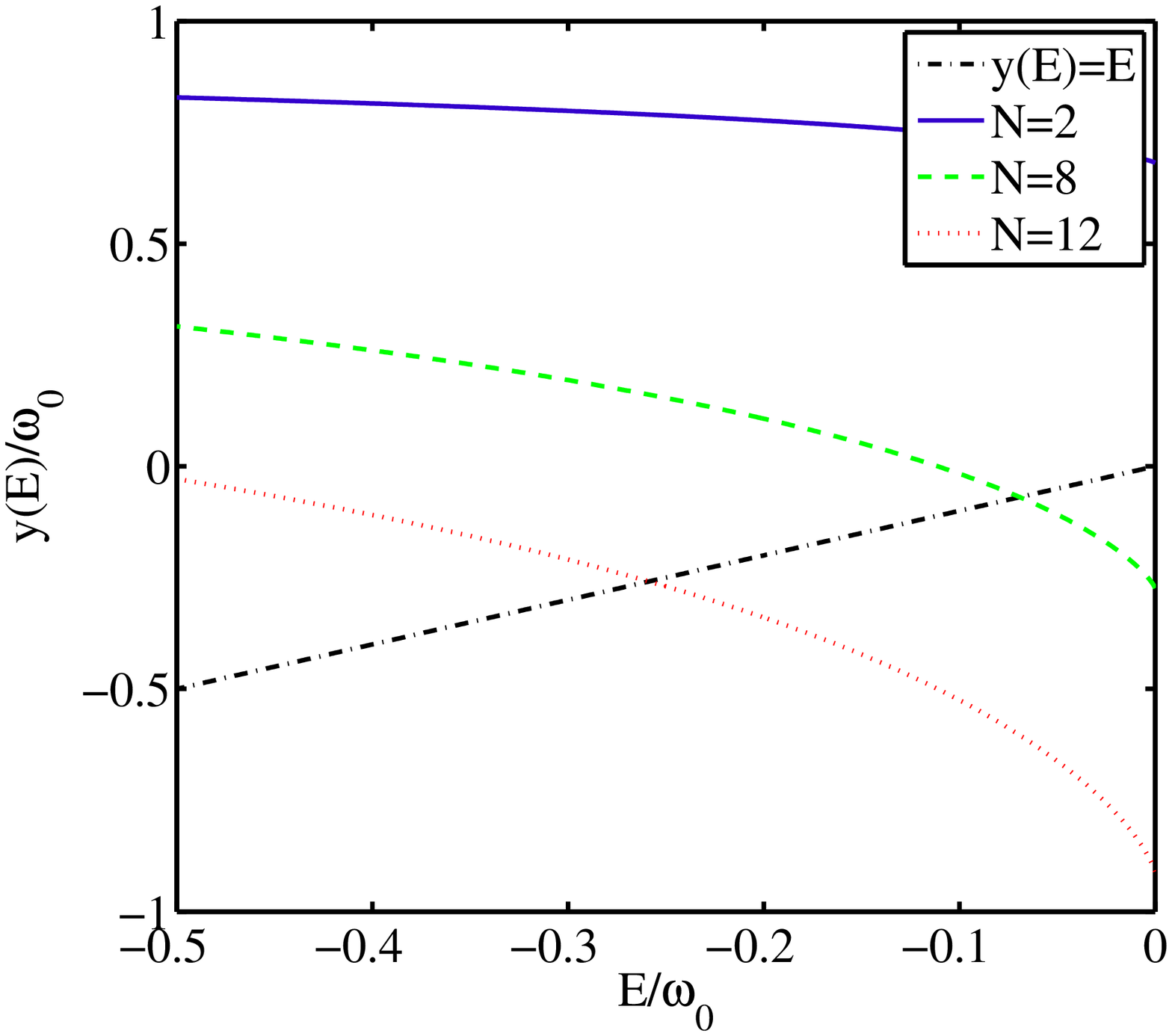}
        \label{fig:first_sub}
    }{
        \includegraphics[width=3in]{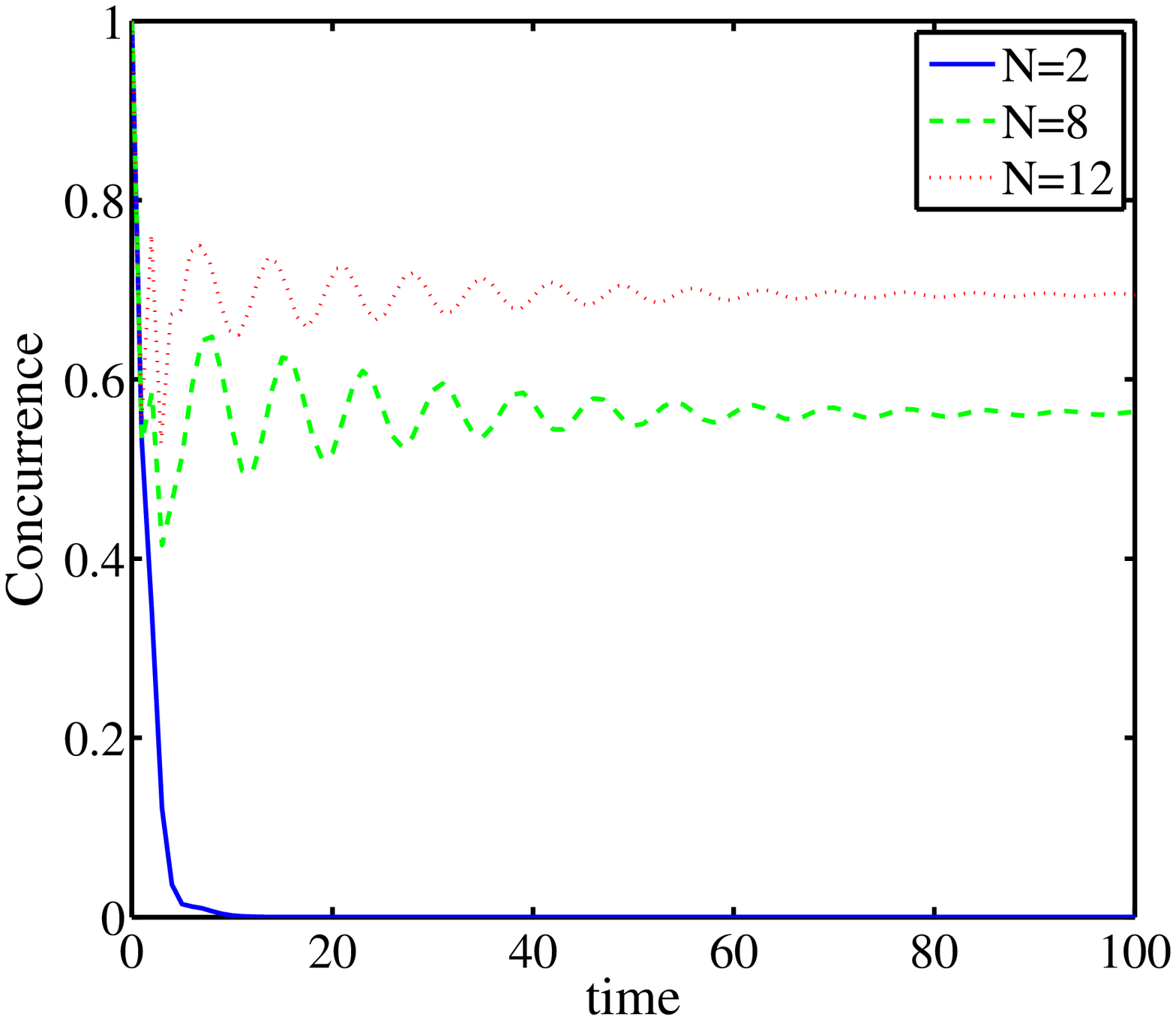}
        \label{fig:second_sub}
    }\\ \par \quad \quad\qquad\qquad\qquad \qquad e \qquad \qquad\qquad\qquad\qquad\quad\qquad\qquad\qquad f\\{
        \includegraphics[width=3in]{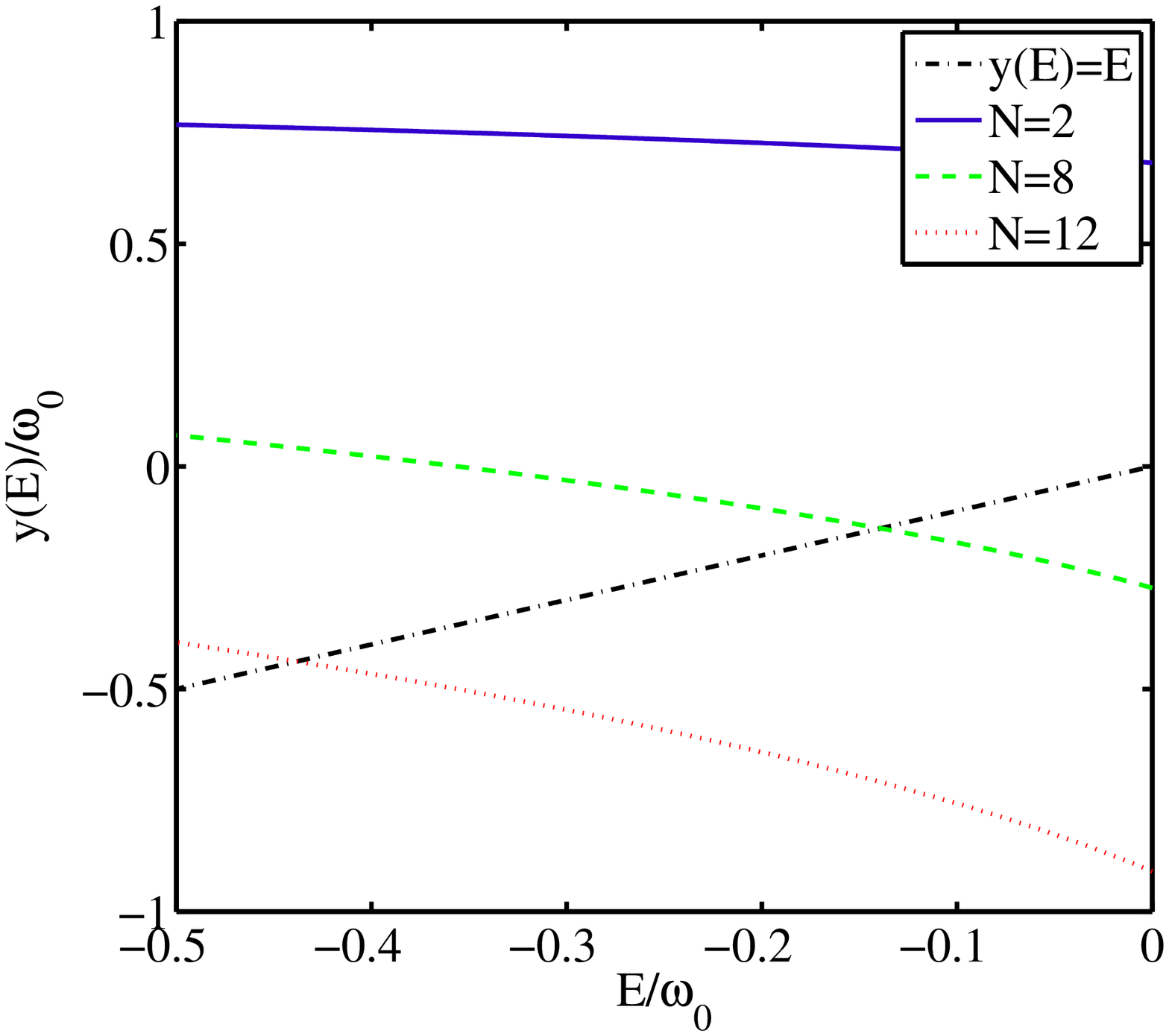}
        \label{fig:first_sub}
    }{
        \includegraphics[width=3in]{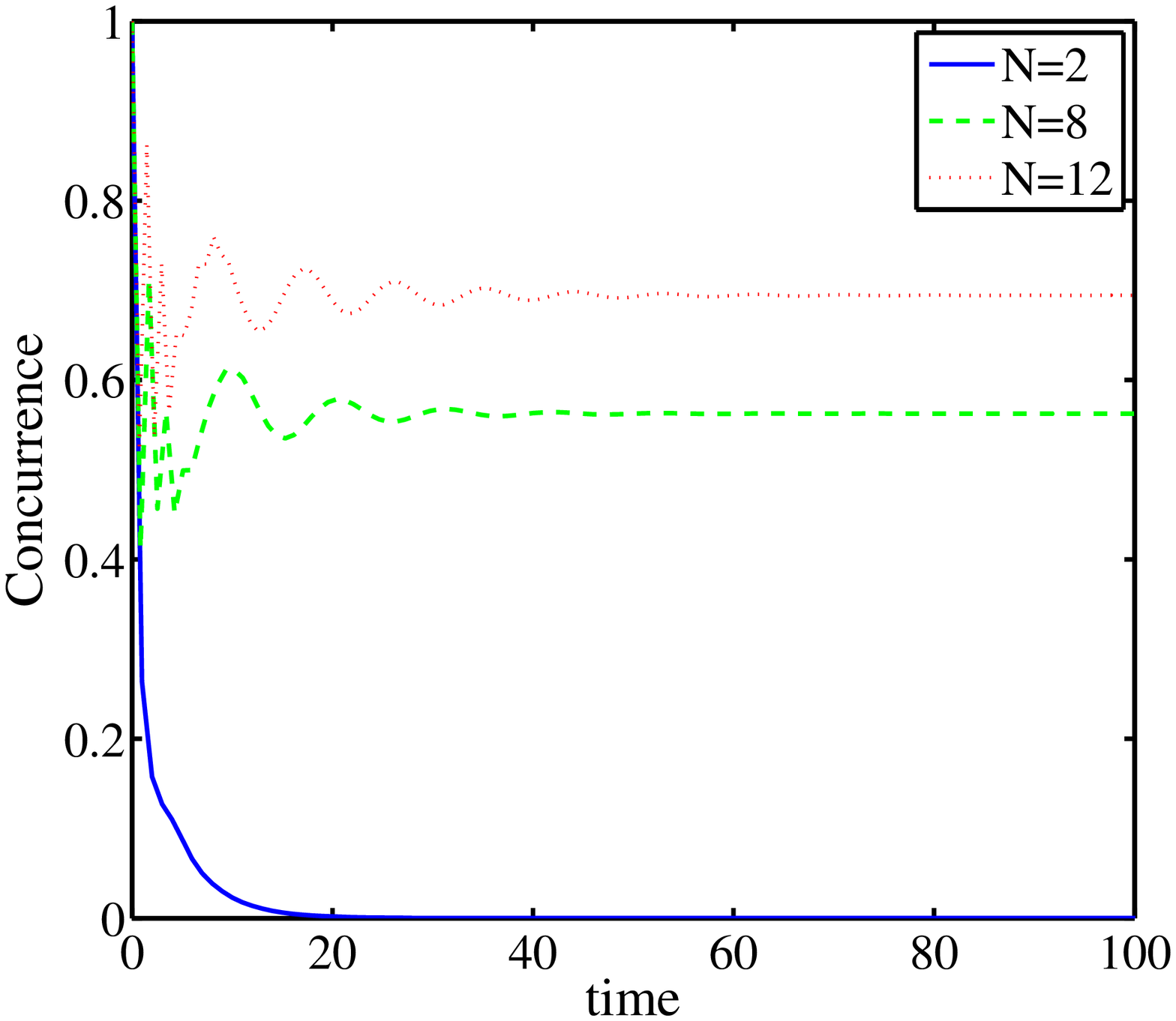}
        \label{fig:second_sub}
    }
    \caption{}
    \end{figure}

\end{document}